\documentclass[fleqn,usenatbib]{mnras}

\usepackage{newtxtext,newtxmath}
\usepackage{amsmath}
\DeclareMathOperator*{\argmin}{arg\,min}

\usepackage[T1]{fontenc}

\DeclareRobustCommand{\VAN}[3]{#2}
\let\VANthebibliography\thebibliography
\def\thebibliography{\DeclareRobustCommand{\VAN}[3]{##3}\VANthebibliography}

\usepackage{graphicx}	
\usepackage{subcaption} 
\usepackage{amsmath,mathtools}	
\usepackage{algorithm}
\usepackage{array}
\usepackage{booktabs} 
\usepackage{tabularx,multirow}
\usepackage{xcolor} 
\usepackage{makecell}
\usepackage[automake,acronym,nohypertypes={acronym}]{glossaries}
\usepackage{footmisc}

 \makeglossaries
\newacronym{coemunet}{COEmuNet}{CO Line Emulator}
\newacronym{rt}{RT}{radiative transfer}
\newacronym{agb}{AGB}{Asymptotic Giant Branch}

\title[COEmuNet]{Emulating CO Line Radiative Transfer with Deep Learning}

\author[S.Su et al.]{
Shiqi Su,$^{1,2}$\thanks{E-mail: ss1421@leicester.ac.uk}
Frederik De Ceuster,$^{3}$
Jaehoon Cha,$^{2}$
Mark I. Wilkinson,$^{1}$
Jeyan Thiyagalingam,$^{2}$\and
Jeremy Yates,$^{4}$
Yi-Hang Zhu,$^{5}$
and Jan Bolte$^{6}$
\\
$^{1}$Department of Physics and Astronomy, University of Leicester, Leicester, LE1 7RH, UK\\
$^{2}$Scientific Computing, Rutherford Appleton Laboratory, Science and Technology Facilities Council, Didcot, OX11 0QX, UK\\
$^{3}$Department of Physics and Astronomy, Institute of Astronomy, KU Leuven, Celestijnenlaan 200D, 3001 Leuven, Belgium\\
$^{4}$Department of Computer Science, University College London, 66-72 Gower Street, London, WC1E 6BT, UK\\
$^{5}$School of Computing and Mathematical Sciences, University of Leicester, Leicester, LE1 7RH, UK\\
$^{6}$Department of Mathematics, Kiel University, Heinrich-Hecht-Platz 6, 24118 Kiel, Germany
}
\date{Accepted XXX. Received YYY; in original form ZZZ}

\pubyear{2024}

\begin{document}
\label{firstpage}
\pagerange{\pageref{firstpage}--\pageref{lastpage}}
\maketitle

\begin{abstract}
Modelling carbon monoxide (CO) line radiation is computationally expensive for traditional numerical solvers, especially when applied to complex, three-dimensional stellar atmospheres. We present COEmuNet, a 3D convolutional neural network (CNN)-based surrogate model that emulates CO line radiation transport with high accuracy and efficiency. It consists of an asymmetric encoder-decoder design that takes 3D hydrodynamical models as inputs and generates synthetic observations of evolved stellar atmospheres. The model is trained on data from hydrodynamic simulations of \gls{agb} stars perturbed by a companion. Given a set of input parameters, including velocity fields, kinetic temperature distribution, and CO molecular number densities, the COEmuNet model emulates spectral line observations with a median relative error of $\sim7\%$ compared to a classical numerical solver of the radiative transfer equation, measured over seven frequency channels and arbitrary viewing directions. Besides, COEmuNet delivers a 1000 times speedup, enabling efficient model fitting to observational datasets, real-time visualization of simulations and progress toward integration in large-scale cosmological simulations.
\end{abstract}



\glsresetall
\section{Introduction}
The atmospheres of \gls{agb} stars are critical to understanding mass loss processes, binary interactions, and the chemical enrichment of the interstellar medium. These extended atmospheres are where mass loss originates, driven by pulsations, radiation pressure on dust, and the gravitational interaction with a companion star. As a result, complex structures, such as spiral patterns, emerge in the stellar winds, making \gls{agb} stars key laboratories for studying how stars contribute to galactic evolution. A key challenge in modelling these systems lies in describing the role of \gls{rt}, which describes how electromagnetic radiation interacts with an emitting and absorbing medium, shaping the emergent spectrum and intensity of stellar winds. In \gls{agb} binaries, solving the RT equation in complex geometries is particularly crucial for understanding how light interacts with the medium and for accurately modelling the observable signatures of these stellar wind structures.

The RT equation is essential for modelling the impact of radiation on gas dynamics through radiation pressure \citep{krumholz2009dynamics}, gas chemistry via photoionization and photodissociation \citep{schmidt1992photoionization}, as well as heating and cooling of gas in localized regions \citep{woitke1996gas}. These processes play a vital role in shaping the observable properties of \gls{agb} winds, enabling predictions that guide our interpretations of these intricate systems.

Observational facilities such as the Atacama Large Millimeter/submillimeter Array (ALMA) and the James Webb Space Telescope (JWST) have significantly enhanced our ability to probe \gls{agb} winds at high spatial resolution. ALMA enables detailed mapping of molecular line emission and dust morphology in AGB envelopes, revealing complex structures such as asymmetric outflows, spiral patterns, and detached shells in binary systems. JWST provides mid-infrared access to dust and molecular features that trace the chemical composition and temperature structure of the inner wind regions. Accurately modeling these observations requires solving the \gls{rt} equation in three dimensions and across broad wavelength ranges. Advanced \gls{rt} modeling is therefore essential not only for interpreting multi-wavelength data from ALMA and JWST, but also for deriving dust formation and binary interactions in \gls{agb} stars.

When using \gls{rt} to model observational data, an efficient numerical solver is essential because repeated computation of the solution can be a computational bottleneck, particularly when exploring large parameter spaces or performing high-resolution simulations. Existing ray-tracing solvers, such as Magritte\footnote{The source code can be found at: \url{https://github.com/Magritte-code/Magritte}} \citep{de_ceuster_magritte_2019, de_ceuster_magritte_2020, de_ceuster_3d_2022,CEULEMANS2024100889}  are primarily designed for solving forward problems. Monte Carlo methods \cite{noebauer2019monte}, while easy to parallelize, are inefficient for media with high optical depth as the trajectories of photon packets are randomly determined by the properties of the medium and photons can therefore become trapped in opaque regions. Although these tools are robust, their reliance on computationally intensive numerical methods limits their efficiency. Moreover, most \gls{rt} solvers, such as RADMC-3D~\citep{dullemond2012radmc3d}, SKIRT~\citep{verstocken2017skirt}, CMacIonize~\citep{vandenbroucke2018monte}, SPHray~\citep{altay2008sphray}, and Lampray~\citep{frostholm2018lampray}, are primarily designed for CPU usage, which limits their ability to fully exploit modern high-performance computational resources like GPUs. As the applications of RT grow in complexity, there is a pressing need for faster, scalable approaches that can reduce computational demands while maintaining required levels of accuracy.

Surrogate models, also known as emulators, provide simplified approximations of more complex, numerical models \citep{wang2014evaluation}. By encoding the behaviour of the underlying complex simulations, surrogate models can offer a computationally efficient alternative while retaining a reasonable precision \citep{thombre2015developing}. The universal approximation theorem \citep{hornik1989multilayer} states that neural networks, particularly feedforward networks, with at least one hidden layer and non-linear activation functions, are capable of arbitrarily accurate approximation to any real-valued continuous function within a compact set. When used as surrogate models, deep neural networks can accurately replicate complex simulation behaviours across varied conditions while significantly reducing computational demands. As a result, neural networks are increasingly being used to accelerate \gls{rt} calculations in a range of situations.

For instance, in the field of remote sensing, \cite{Le2020} applied multi-level perceptron to fast hyperspectral \gls{rt} simulations, achieving two to three orders of magnitude in computational speed-up while maintaining small relative errors. In astronomy, \cite{sethuram2023emulating} introduced ANNgelina to emulate computationally expensive Monte Carlo \gls{rt} calculations for galaxy spectral energy distributions. EmulART, proposed by \cite{rino2023emulart}, uses a denoising variational autoencoder for spectral compression, combined with an approximate Bayesian approach for spatial inference, to efficiently recreate high-resolution \gls{rt} models from low-information inputs.
\cite{lu2024surrogate} discusses the use of physics-informed neural operators to develop surrogate models for simulating radiative heat transfer. \cite{mishra2021physics} introduces a physics informed neural network (PINN) framework for solving the \gls{rt} equation by embedding physics constraints into a neural network to predict the intensity field, in both forward and inverse problems. Despite these advancements demonstrating the success of deep learning-based surrogate models for \gls{rt} modelling, surrogate models specifically tailored for spectral line radiative transfer to model position-velocity data in astrophysical contexts remain underexplored. Developing such a model would benefit astrophysical studies by significantly reducing computation times and resource demands while maintaining required levels of accuracy, opening up the possibility of including realistic \gls{rt} calculations in larger numerical simulations for which \gls{rt} effects would normally be incorporated using simplified "sub-grid" physics models~\citep{teyssier2019numerical}, and making tractable the interpretation of complex observed structures as observed in the mid-infrared by the JWST and the sub-mm/mm by ALMA. A deep learning-based surrogate model would also leverage modern accelerators, such as GPUs, to significantly enhance computational efficiency.

In this paper, we propose a three-dimensional convolutional neural network-based surrogate model, \gls{coemunet}, to efficiently generate synthetic position-velocity observations (RA,Dec and vz) of hydrodynamic models of \gls{agb} stellar winds. \gls{coemunet} is designed to handle multiple frequencies (\(\nu\)) and image an object from any viewing angle (\(\boldsymbol{\hat{n}}\)), offering a computationally efficient alternative to traditional \gls{rt} solvers. Our approach was trained and validated through comparisons with Pomme\footnote{The Pomme source code can be found at: \url{https://github.com/Magritte-code/pomme}}~\citep{de2024bayesian}, utilizing Pomme's simplified forward \gls{rt} equation line solver. The model not only fills the gap in \gls{rt} modelling for spectral line imaging in astrophysics, but also demonstrates substantial reductions in computation time while maintaining sufficient accuracy to reproduce observational images.

The paper is organised as follows. Section \ref{sec:methodology} introduces the simulated datasets used to train the \gls{coemunet}, followed by the architecture of \acrshort{coemunet} and the training setup. Section \ref{sec:results} discusses the performance of \acrshort{coemunet} and Section \ref{sec:conclusion} concludes the paper.

\section{Methodology}\label{sec:methodology}

\subsection{Forward Radiative Transfer}
This initial study focuses on modelling the rotational transition line of carbon monoxide (CO) from quantum number $J=1$ to $J=2$. This choice is motivated by CO's high abundance and ease of excitation, making it readily observable. For simplicity, scattering effects are excluded as they are negligible at these wavelengths. To validate the surrogate model, local thermodynamic equilibrium (LTE) is assumed, ensuring consistency with the assumptions of the Pomme solver \citep{de2024bayesian}, which serves as the reference for comparing the \gls{coemunet} model output. However, we emphasise that we have not explicitly used the LTE assumption in constructing our surrogate model. We also note that while LTE is not the correct assumption for modelling circumstellar CO emission, for the purpose of this proof of concept paper, we make the LTE assumption simply due to the greater availability of LTE-based training data. The extension of our surrogate model to non-LTE data would be straightforward given the  availability of sufficient non-LTE data.

In this paper, we consider the time-independent \gls{rt} equation excluding scattering effects, given by,
\begin{equation}
\boldsymbol{\hat{n}} \cdot \nabla I_{\nu}(\boldsymbol{x},\boldsymbol{\hat{n}})  = \eta_{\nu}(\boldsymbol{x}) - \chi_{\nu}(\boldsymbol{x}) I_{\nu}(\boldsymbol{x},\boldsymbol{\hat{n}})
\label{eq:radiative transfer}
\end{equation}
where the specific intensity of the radiation field, $I_{\nu}(\boldsymbol{x},\boldsymbol{\hat{n}})$, is a function of frequency $\nu$, position $\boldsymbol{x}$ and direction $\boldsymbol{\hat{n}}$, 
and $\eta_{\nu}(\boldsymbol{x})$ and $\chi_{\nu}(\boldsymbol{x})$ are the emissivity and opacity of the medium, respectively, as functions of frequency and position. Equation~\ref{eq:radiative transfer} describes the change of specific intensity along the line of sight $\boldsymbol{\hat{n}}$.

The formal solution to Equation~\ref{eq:radiative transfer} is 
\begin{equation}
\begin{split}
I_{\text{obs}}(\nu; x, y)
\ =& \ \,
I_{\text{bdy}}(\nu; x, y) \, e^{-\tau_{\text{obs}}(\nu; \, x, y, L)} \\
& \ + \
\int_{0}^{L} \eta\big( \nu_{\text{com}}(\nu; \, x, y, z); \, x, y, z \big) \, e^{-\tau_{\text{obs}}(\nu; \, x, y, z)} \text{d}z .
\label{eq:formalsolution}
\end{split}
\end{equation}
Here $I_{\text{obs}}$ is the observed intensity at frequency $\nu$ in the observer frame calculated along the $z$-axis assuming a calculation box of length $L$, $\nu_{\text{com}}$ is the frequency in the co-moving frame of a fluid element of the medium and $I_{\text{bdy}}(\nu; x, y)$ is the intensity of the incoming radiation at the boundary of the model box at $(x,y,0)$. (see Figure~\ref{fig:hydro_grid} for a schematic representation of the grid structure). Finally, the optical depth along the $z-$axis is defined as
\begin{equation}
	\tau_{\text{obs}}(\nu; x, y, z)
	\ \equiv \
	\int_{0}^{z} \chi\big( \nu_{\text{com}}(\nu; \, x, y, z'); \, x, y, z' \big) \text{d}z'.
	\label{eq:opticaldepth}
\end{equation}
To compute $I_{\text{obs}}$, Equation~\ref{eq:formalsolution} requires the monochromatic line emissivity, $\eta_{\nu}(\boldsymbol{x})$, and opacity, $\chi_{\nu}(\boldsymbol{x})$ associated with the spectral line transition between the quantized energy levels $i=2$ and $j=1$. These are expressed as:
\begin{align}
\eta_{ij}(\nu,x) \ &= \  \eta_{ij}(x) \, n(x) \, \phi_{ij}(\nu, x) , \label{eq:emi} \\
\chi_{ij}(\nu,x) \ &= \  \chi_{ij}(x) \, n(x) \, \phi_{ij}(\nu, x) . \label{eq:opa}
\end{align}
where $n(\boldsymbol{x})$ represents the number density of the chemical species responsible for the line, while $\phi_{ij}(\nu,\boldsymbol{x})$ is the line profile function that describes the spectral distribution of the line.

In this paper, we will assume the Gaussian line profile function $\phi_{ij}(\nu,\boldsymbol{x})$  predominantly caused by Doppler shifts due to the thermal and turbulent motions (along the line of sight) in the medium. The line profile is centered around the line frequency, $\nu_{ij}$, and is expressed as,
\begin{equation} 
\phi_{ij}(\nu,\boldsymbol{x}) \ =
\frac{1}{\delta\nu_{ij}(\boldsymbol{x}) \sqrt{\pi}} \ \exp \left( - \left(\frac{\nu - \nu_{ij}}{\delta \nu_{ij}(\boldsymbol{x})}\right)^{2} \right), 
\end{equation}
in which the line width, $\delta\nu_{ij}(\boldsymbol{x})$ is defined through the local temperature $T(\boldsymbol{x})$, the molecular mass $m_{spec}$ and a local turbulent velocity $v_{\text{turb}}$, 
\begin{equation} \delta\nu_{ij}(\boldsymbol{x}) \ =
\frac{\nu_{ij}}{c} \sqrt{\frac{2 k_{\text{B}}T(\boldsymbol{x})}{m_{\text{spec}}} + v_{\text{turb}}^{2}(\boldsymbol{x})}. 
\end{equation}
The two remaining components $\eta_{ij}(\boldsymbol{x})$ and $\chi_{ij}(\boldsymbol{x})$ in Equation \ref{eq:emi} and \ref{eq:opa} are the line emissivity and opacity defined as: 
\begin{align}
\eta_{ij}(\boldsymbol{x}) \ &= \  \frac{h \nu_{ij}}{4 \pi} \, p_{i}(\boldsymbol{x}) A_{ij}, \\
\chi_{ij}(\boldsymbol{x}) \ &= \  \frac{h \nu_{ij}}{4 \pi} \, \Big( p_{j}(\boldsymbol{x}) B_{ji} - p_{i}(x) B_{ij} \Big) ,
\end{align}
where $A_{ij}$ represents the Einstein coefficient for spontaneous emission, $B_{ij}$ is the coefficient for stimulated emission, and $B_{ji}$ is the coefficient for absorption.

Under the LTE assumption, the population of energy levels $p_{i}(\boldsymbol{x})$ is governed by the local gas kinetic temperature, $T(x)$, such that:
\begin{equation}
    p_{i}(\boldsymbol{x})
    \ = \
    \frac{g_{i}}{Z(\boldsymbol{x})} \,
    \exp \left(-\frac{E_{i}}{k_{\text{B}} T(\boldsymbol{x})} \right) ,
\end{equation}
where $g_{i}$ and $E_{i}$ denote the statistical weight and the energy of the level $i$, $k_{\text{B}}$ is the Boltzmann constant, $Z(\boldsymbol{x})$ is the partition function, ensuring the level populations are properly normalised ($\sum p_{i}(\boldsymbol{x})=1$).
To account for the Doppler effect caused by the motion of the medium along the line of sight, we shift the frequencies to the observer's frame,  
\begin{equation}
    \nu_{\text{obs}}(\nu, \boldsymbol{x}) \ = \ \left( 1 + \frac{v_{z}(\boldsymbol{x})}{c} \right) \nu ,
\label{eq:doppler_shift}
\end{equation}  
where $v_{z}(\boldsymbol{x})$ is the $z$-component of the velocity field and $c$ is the speed of light.

All other parameters, such as the radiative constants, can be found in databases. In this paper, we use the Leiden Atomic and Molecular Database\footnote{\label{footnote:lamada}The database can be found at: \url{https://home.strw.leidenuniv.nl/~moldata/.}} \citep[LAMDA;][]{schoier_atomic_2005}.

\subsection{Data preparation}
\subsubsection{Dataset}\label{sec:train_data}
The main goal of this study is to develop an efficient surrogate model to construct synthetic observations of hydrodynamic models of complex stellar wind structures around \gls{agb} stars in binaries. We therefore require a training dataset consisting of 3D hydrodynamical models and the corresponding synthetic observations. Clearly, constructing such a dataset with real observations is impossible, as it would require complete knowledge of the physical properties throughout the entire wind structure. We do not have a ground truth dataset derived from observational data. Instead, we generate our training dataset using numerical simulations.

The hydrodynamical models were generated using the Adaptive Mesh Refinement Versatile Advection Code \citep[AMRVAC;][]{keppens2003adaptive,xia2018mpi,keppens2023mpi}. These simulations model the stellar winds of a central mass-losing \gls{agb} star, whose atmosphere is perturbed by a companion star. To produce synthetic observations from these models, we solve the RT equations numerically using the \gls{rt} solver in Pomme~\citep{de2024bayesian}. The hydrodynamical simulations were specifically designed to interpret observations from the ATOMIUM ALMA Large Program~\citep{decin2020sub,gottlieb2022atomium}.

Our dataset comprises pairs of hydrodynamic models and corresponding synthetic images. Multiple data pairs are generated from a single hydrodynamic model by imaging it from different viewing angles (see implementation details in Appendix~\ref{sec:rotate_hydro_model}). The full line profile is approximated by computing images in seven frequency bins (\(n_\nu=7\)), each of width $h=0.000230698$ GHz centred on $\nu_{0}=230.5380$ GHz \footref{footnote:lamada}, the peak of the line profile. This frequency range was chosen to balance capturing all relevant spectral features including Doppler shifts (see Equation \ref{eq:doppler_shift}) in the spectral lines while minimizing the size of the training dataset. The selected seven frequency bins encompass the richest spectral features, whereas frequencies outside this range primarily consist of background noise.

The complete dataset is defined as:
\[
\mathcal{D} = \{(\boldsymbol{\xi}_i, I_\nu(\boldsymbol{x}, \boldsymbol{\hat{n}}_i))\}_{1 \leq i \leq N},
\]
where \(N=2180600\) is the number of samples. This is derived from a set of 10903 hydrodynamic models, each used to generate 200 data paris. Each sample consists of the input \(\boldsymbol{\xi}_i = \{v_z(\boldsymbol{x}), T(\boldsymbol{x}), n_{CO}(\boldsymbol{x}) \mid \boldsymbol{x} = (x, y, z)\}\), which includes three position-dependent physical parameters: velocity \(v_z(\boldsymbol{x}) \in \mathbb{R}^{N_x^{H} \times N_y^{H} \times N_z^{H}} \); kinetic temperature, \(T(\boldsymbol{x}) \in \mathbb{R}^{N_x^{H} \times N_y^{H} \times N_z^{H}} \); and  CO number density, \(n_{CO}(\boldsymbol{x}) \in \mathbb{R}^{N_x^{H} \times N_y^{H} \times N_z^{H}}\). The spatial domain of the hydrodynamic model $H$ is represented on a Cartesian grid with resolution $N_x^{H} = N_y^{H} = N_z^{H} = 64$, chosen to balance computational cost while adequately resolving the relevant structures. Notably, only the \(z-\) component of velocity, \(v_z(\boldsymbol{x})\), is used, as the numerical solver calculates the intensity \(I_{\nu}\) along the \(z-\)axis. 

The target intensity images, $I_\nu(\boldsymbol{x},\boldsymbol{\hat{n}}_i) \in \mathbb{R}^{n_\nu \times N_x^{I} \times N_y^{I}}$, are computed at seven frequency bins ($n_\nu=7$) along the direction $\boldsymbol{\hat{n}}_{i}$.  Here,  $N_x^{I}$ and $N_y^{I}$ denote the number of spatial bins in the synthetic image $I$ along the $x-$ and $y-$axes.

The dataset \(\mathcal{D}\) is randomly split into 80\% training, 10\% validation and 10\% test sets.
\subsubsection{Data Preprocessing}
In this study, distinct preprocessing strategies are applied to the data on the three physical quantities in the input dataset: $v_z(\boldsymbol{x})$, $T(\boldsymbol{x})$), and $n_{CO}(\boldsymbol{x})$. The velocity component, $v_z(\boldsymbol{x})$ is standardized using 
\begin{equation}
    v_z'(\boldsymbol{x}) = \frac{v_z(\boldsymbol{x}) - \mu}{\sigma}
\end{equation}
where $\mu$ and $\sigma$ are the mean and standard deviation of $v_z(\boldsymbol{x})$, respectively. \\

The kinetic temperature distribution \(T(\boldsymbol{x})\), in the hydrodynamic models is highly skewed and spans a wide dynamic range. To address this, a logarithmic transformation is applied, followed by normalization using the median value, as the median is more robust to skewed distribution than the mean. The preprocessing is defined as: 
\begin{align} 
t(\boldsymbol{x}) &= \ln(T(\boldsymbol{x})), \\
\ T'(\boldsymbol{x}) &= \frac{t(\boldsymbol{x}) - \min(t)}{\operatorname{median} \left( t(\boldsymbol{x}) - \min(t) \right)}, \label{eq:preprocess} 
\end{align} 
where $t(\boldsymbol{x})$ is log-transformed temperature, and $T'(\boldsymbol{x})$ is the normalized value. Here, $\textit{min}(\cdot)$ and $\textit{median}(\cdot)$ return the minimum and median values of their respective arguments. The same transformation is applied to the CO number density, $n_{CO}(\boldsymbol{x})$.

For the target intensity images, $I_{\nu}(\boldsymbol{\xi}_i,\boldsymbol{\hat{n}}_i)$, a  logarithmic transformation is applied followed by min-max normalization.
\begin{align} 
 t_{\nu} &= \ln(I_{\nu}(\boldsymbol{\xi}_i,\boldsymbol{\hat{n}}_i)), \\
 I'_{\nu}(\boldsymbol{\xi}_i,\boldsymbol{\hat{n}}_i) &= \frac{t_{\nu} - \min(t_{\nu})}{\max(t_{\nu}) - \min(t_{\nu})}.
\end{align} 
The logarithmic transformation compresses the dynamic range of the data, while min-max normalization scales the values to the range $[0,1]$.

\subsection{AI preliminaries}

\subsubsection{Convolutional Neural Network}
Convolutional neural networks (CNNs) \citep{cnn1,cnn2} are widely used for processing grid-structured data, such as images and 3D volume data. A 3D convolution extends the concept of 2D convolution by incorporating a depth dimension alongside height and width, enabling the analysis of volumetric structures. CNNs take feature maps as input, which encodes structured representation over a discrete domain $\mathbb{Z}^{n}$, and convolve them with a set of $c_{out}$ filters (kernels). Formally, a feature map is defined as a function $f:\mathbb{Z}^{n} \rightarrow\mathbb{R}^{c_{in}}$, where $c_{in}$ denotes the number of input channels. In the present case, we have three feature maps: the $v_z(\boldsymbol{x})$, $T(\boldsymbol{x})$ and $n_{CO}(\boldsymbol{x})$ data cubes. Each filter $\gamma^{i}$ in the set $\Gamma = \{\gamma^{l}\}^{c_{out}}_{i=1}$ is a function $\gamma^{i}:\mathbb{Z}^{n} \rightarrow\mathbb{R}^{c_{in}}$. The convolution operation is defined as:
\begin{equation} 
(f \star \gamma^{i})(\boldsymbol{u}) = \sum_{\boldsymbol{y} \in \mathbb{Z}^{n}} \langle f(\boldsymbol{y}), \gamma^{i}(\boldsymbol{y} - \boldsymbol{u}) \rangle, 
\end{equation}
which can be interpreted as an inner product of input feature maps with the corresponding filters at every point $y\in\mathbb{Z}^{n}$. The filter size of a convolution layer is a crucial choice in neural network design since it defines the regions from which information is obtained. Common practice is to use rather small filters \citep{simonyan2014very}. 
\subsubsection{Batch Normalization}\label{sec:batch_norm}
Batch normalization (BN)~\citep{ioffe2015batch} stabilizes training by normalizing feature values within each mini-batch (a subset of dataset) and preventing large fluctuations in gradient updates. For a layer of the network with $d-$dimensional input and mini-batch $B$ of size $m$, $\boldsymbol{u}=(\boldsymbol{u}^{(1)},...,\boldsymbol{u}^{(d)})$, each dimension of its input is then normalised, 
\begin{align}
    \hat{u}^{(k)}_i &= \frac{u^{(k)}_i - \mu^{(k)}_B}{\sqrt{(\sigma^{(k)}_B)^2+\epsilon}}, k \in [1,d], i \in [1,m]\\
    \mu_B^{(k)} &= \frac{1}{m} \sum_{i-1}^{m}u_i^{k}, \quad \ \sigma^{(k)}_{B} = \frac{1}{m} \sum_{i-1}^{m}(u_i^{k}-\mu_B^{k})^2\\
\end{align}
where \(\mu^{(k)}_B\) and \(\sigma ^{(k)}_B\) are the mean and standard deviation. $\epsilon$ is added in the denominator for numerical stability and is an arbitrarily small constant. The normalized feature $\hat{u}^{(k)}_i$ is then scaled and shifted using learnable parameters \(\alpha,\beta\) to give output \(y^{(k)}_i\):
\begin{equation}
    y^{(k)}_i = \alpha ^{(k)} \ \hat{u}^{(k)}_i + \beta^{(k)}_i
\end{equation}

\subsubsection{Residual Blocks and Skip Connections}
A residual block \citep{he2016deep} consists of a sequence of convolutional layers, each followed by batch normalization and a non-linear activation (typically rectified linear unit (ReLU)~\citep{agarap2018deep}).  Residual blocks were introduced to address the vanishing gradient problem in deep neural networks, facilitating the training of very deep architectures. Instead of directly learning the desired mapping \(\boldsymbol{y}(\boldsymbol{u})\), the block learns a residual function $\mathcal{F}$ that represents the difference between \(\boldsymbol{y}(\boldsymbol{u})\) and the input \(\boldsymbol{u}\). Formally, this relationship is expressed as:
\[\boldsymbol{y}(\boldsymbol{u}) = \mathcal{F}(\boldsymbol{u},\{W_i\}) + \boldsymbol{u}\]
where $\{W_i\}$ denotes the set of learnable weights within the residual block. If the optimal mapping is close to the identity, the residual function $\mathcal{F}$ learns to approximate zero, allowing the input $\boldsymbol{u}$ to pass through unchanged. As a result, when updating the gradients of the network's parameters, gradients propagate directly through the shortcut connection, significantly reducing the risk of vanishing gradients.

Skip connections are a key element in residual blocks. When the dimensions of the input and output are the same, an identical skip connection is used. Input is directly added to the output of residual function:
\begin{equation}
    \boldsymbol{y} = \mathcal{F}(\boldsymbol{u},\{W_i\}) + \boldsymbol{u}
\label{eq:residual_identity}
\end{equation}
If dimensions differ, a residual connection is usually implemented as convolution layer with kernel size 1 (a learnable weight \(W_s\)):
\begin{equation}
    \boldsymbol{y} = \mathcal{F}(\boldsymbol{u},\{W_i\}) + W_s \boldsymbol{u}
\label{eq:residual}
\end{equation}

\subsubsection{Standard paradigm of deep learning}
Training a deep learning model involves iteratively updating its parameters to minimize the discrepancy between the model's prediction and the underlying target, given a dataset \(\mathcal{D} = \{(q_i, p_i)\}_{i=1}^{N}\), where \(q_i\) represents the input and \(p_i\) the corresponding target. It is partitioned into a training set \(\mathcal{D}_{\text{train}}\) and a validation set \(\mathcal{D}_{\text{val}}\). The model is trained on \(\mathcal{D}_{\text{train}}\), while \(\mathcal{D}_{\text{val}}\) is used to assess generalization performance through predefined evaluation metrics (quantitative measures). 

The network learns a function \(f_\theta(q)\) parameterized by \(\theta\) to minimize a loss function \(\mathcal{L}(f_\theta(q), p)\). The outcome is a set of optimal parameters $\theta^{*}$ satisfying
\begin{equation*}
    \theta^{*} \in \Theta: \theta^{*} = \argmin_{\theta^{*} \in \Theta} \mathcal{L}(f_\theta(q), p)
\end{equation*}
Training proceeds over multiple epochs, where each epoch corresponds to one complete pass over \(\mathcal{D}_{\text{train}}\). During each iteration, $\theta$ is updated using gradient-based optimization: 
\begin{equation}
    \theta_{t+1} = \theta_t - \eta \nabla_\theta \mathcal{L}
\end{equation}
where \(\eta\) is the learning rate, a crucial hyperparameter that controls the step size of updates. Selecting an appropriate \(\eta\) is essential, as values that are too large may lead to divergence, while values that are too small can slow convergence. A grid search is conducted over multiple learning rate candidates, selecting the value that achieves the best model performance on \(\mathcal{D}_{\text{val}}\). To improve training efficiency, a learning rate scheduler dynamically adjusts \(\eta\) over epochs. A common choice is the StepLR scheduler, which updates \(\eta\) as follows:
\begin{equation}
    \eta_t = \delta \eta_{t - s}
\end{equation}
where \(\delta\) is the decay factor and \(s\) is the step size in epochs.

Following this formulation, 
the surrogate model \( \tilde{I}_{\theta}(\boldsymbol{\xi}_i)\) is trained with parameters $\theta$ to emulate the target output \(I_\nu(\boldsymbol{\xi}_i, \boldsymbol{\hat{n}}_i)\).

\subsection{Surrogate Model Architecture}\label{sec:method}
\begin{figure*}
    \centering
    \includegraphics[width=\linewidth]{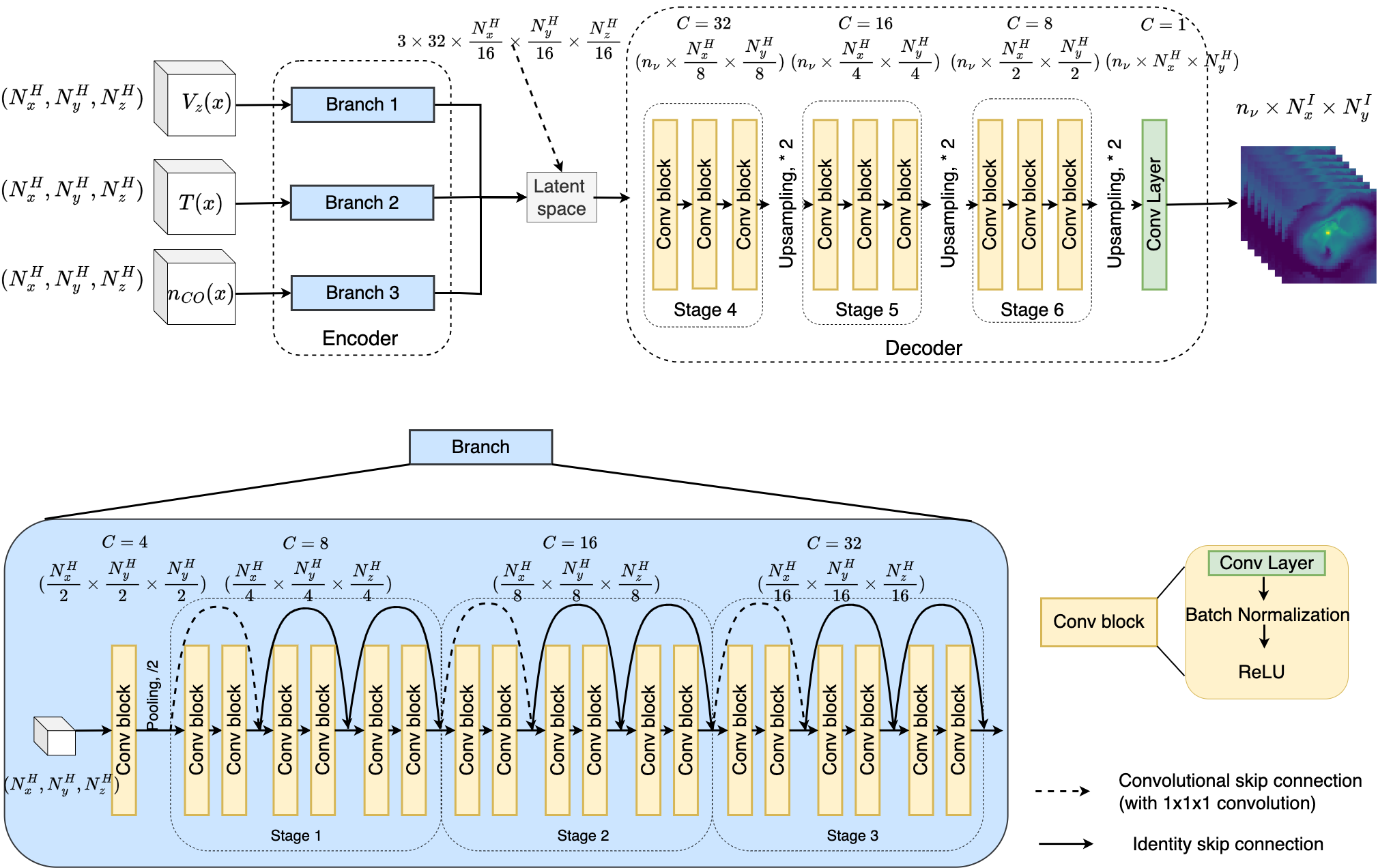}
    \caption{Overview of the \acrshort{coemunet} architecture. The model consists of an encoder and a decoder. The encoder has three separate branches, each processing a specific spectral feature: $V_z{(\boldsymbol{x})}, T(\boldsymbol{x}), n_{CO}(\boldsymbol{x})$). Each 
    branch applies convolutional blocks with residual connections, progressively increasing feature channels while reducing spatial resolution. The latent space concatenates the encoded features and refines them through fully connected layers before reshaping them into a 3D tensor. The decoder then reconstructs the output by applying a series of $1\times 3 \times 3$ convolutional layers and nearest-neighbor upsampling. Solid arrows denote identity skip connections and dashed arrows denote convolution skip connections.}
    \label{fig:resnet}
\end{figure*}
The \acrshort{coemunet} architecture consists of an encoder and a decoder, as illustrated in Fig.~\ref{fig:resnet}. COEmuNet is specifically designed for settings where the input consists of physical quantities and the output is an image, with no structural similarity between them—unlike architectures such as U-Net that rely on shared spatial features. The encoder extracts feature representations from the input parameters, the latent space compresses and encodes these representations in a lower-dimensional form, and the decoder reconstructs the output from the latent representation.

The encoder has three separate branches, each applied to a specific physical quantity: $v_z(\boldsymbol{x}),T(\boldsymbol{x}) \text{ and } C(\boldsymbol{x})$. By keeping these quantities separate, the model can learn distinct features from each while reducing interference between them. Each branch consists of residual blocks with a kernel size 3, progressively increasing the number of feature channels from 8 to 16 to 32 while simultaneously reducing the spatial resolution. The spatial dimensions are compressed to resolutions of 16, 8, and 4 in stages 1, 2, and 3, respectively. 

In the latent space, the encoded features from all three branches are concatenated into a flattened tensor, which is then refined by two fully connected layers. 

The decoder reconstructs the output using a series of 3D convolutions while progressively increasing spatial resolution through nearest-neighbour interpolation. The number of feature channels and spatial resolution in stages 4, 5, and 6 are set to (32, 16, 8) and (8, 16, 32), respectively. Although upsampling restores the spatial dimensions, a final convolutional layer is applied to refine the interpolated features, ensuring that learnable kernels enhance the feature representations.
\subsection{Training setup}
We implemented \acrshort{coemunet} using the PyTorch deep learning framework \citep{paszke2019pytorch}. To ensure reproducibility, a fixed random seed is set, guaranteeing that the trained weights remain consistent across multiple executions of the algorithm.
\subsubsection{Data Management Strategies}
As a 7TB dataset cannot fit entirely into memory on a typical compute node in an HPC cluster, preprocessing the entire dataset at once is impractical. To address this, the dataset is divided into smaller, manageable chunks, and local statistics are computed independently for each chunk. These statistics include the mean ($\mu_i$), variance($\sigma_{i}^{2}$), median ($M_i$), and minimum and maximum values.The global statistics for the entire dataset are then obtained by aggregating these local statistics. Specifically, the global mean and variance ($\mu,\sigma^2$) are computed using a parallel algorithm~\citep{chan1982updating}, while the global median ($M$) is determined using the median of medians algorithm \citep{blum1973time}. Similarly, the global minimum and maximum values are simply the minimum and maximum of the local minima and maxima, respectively.

During training, an I/O bottleneck was observed due to frequent disk access when loading individual samples into CPU RAM, significantly slowing model progress. To mitigate this issue, data chunks are preloaded into CPU memory in advance, reducing the frequency of disk access. Since the CPU and GPU reside on the same node and are directly connected, PyTorch can efficiently preload the next chunk of data in parallel using multiple threads while the GPU processes the current batch. This approach ensures that batches are readily available in memory, minimizing GPU idle time and maintaining a smooth and efficient data flow throughout the training process.

\subsubsection{Pretraining}
The \acrshort{coemunet} was pre-trained on a simplified dataset to establish a robust initialization. Based on an initial test, pretraining facilitates faster convergence and improves initialization for subsequent fine-tuning. The pretraining dataset consisted of intensity $I_{\nu}(\boldsymbol{\xi},\boldsymbol{\hat{n}})$ observed exclusively in a face-on orientation (i.e., along the axis perpendicular to the orbital plane of the binary) at a single frequency ($n_\nu = 1$). Training was conducted over 1000 epochs using the Adam optimizer \citep{kingma2014adam} with a learning rate of 0.004, regulated by a StepLR learning rate scheduler. Pretraining was performed on four A100-SXM-80GB GPUs, with each GPU processing a batch of 128 samples, resulting in a total batch size of 512. It required approximately 8 GPU hours (four GPUs for 2 hours of wall-clock time). The detailed training configuration is provided in Table~\ref{tab:faceon_pretrain}.

\subsubsection{Fine-tuning}
Following pretraining, the \acrshort{coemunet} model was fine-tuned on the complete dataset $\mathcal{D}$, which included intensities observed across seven frequencies ($n_\nu=7$) and arbitrary viewing directions \(\hat{n}=(\theta,\phi)\). Specific adjustments (Table~\ref{tab:mulfreq_finetune}) were implemented to capture the intricate patterns present in $\mathcal{D}$. Fine-tuning was performed on four A100-SXM-80GB GPUs, with each GPU processing a batch of 512 input-output pairs, resulting in a total batch size of 2048. The Adam optimizer was retained with a fixed learning rate of 0.0001. This phase enabled the model to generalize effectively to the full range of frequencies and observing directions while building on the foundational knowledge gained during pretraining. The fine-tuning required approximately 288 GPU hours, corresponding to 72 hours of wall-clock time on four A100-SXM-80GB GPUs.

\subsubsection{Loss function}
The network was trained using a loss function that combines mean squared error (MSE) and a frequency-based loss:
\begin{equation}
    \mathcal{L} = \lambda \mathcal{L}_{\text{MSE}} + (1-\lambda) \mathcal{L}_{\text{freq}}
\end{equation}
where $\lambda$ is a weighting hyperparameter that balances the contributions of $\mathcal{L}_{MSE}$ and $\mathcal{L}_{freq}$. We select \(\lambda\) from a predefined set of values using a grid search and choose the value that maximizes model performance based on validation metrics.\\
The MSE quantifies the pixel-wise difference between target image, \(I_{\nu}(x,y)\), and emulated image,  \(I'_{\nu}(x,y)\),
\begin{equation}
    \mathcal{L}_{MSE} = 
    \frac{1}{n_{\nu} N_{x}^{I} N_{y}^{I}} \sum_{\nu=1}^{n_{\nu}} \sum_{x=1}^{N_{x}^{I}} \sum_{y=1}^{N_{y}^{I}} (I_{\nu}(x,y) - I'_{\nu}(x,y))^2
\label{eq:MSE}
\end{equation}
where $N_{x}^{I}$ and $N_{y}^{I}$ denote the image width and height, respectively, and $(x, y)$ is the spatial coordinate of a pixel.\\
Unlike pixel-wise losses, frequency loss explicitly operates in the frequency domain, measuring discrepancies between predicted and ground truth images based on their frequency components. This approach is more sensitive to structural details such as textures and edges and can prioritize specific frequency ranges to emphasize particular image characteristics. To enhance sharpness and texture reconstruction, we introduce a frequency loss function \citep{zhang2020single}, defined as:
\begin{equation}
     L_{\text{freq}}  = \frac{1}{n_{\nu} N_{x}^{I} N_{y}^{I}}  \sum_{\nu=1}^{n_{\nu}} \sum_{u=1}^{N_{x}^{I}} \sum_{v=1}^{N_{y}^{I}} |F(u,v) - F'(u,v)| \label{eq:l1f}
\end{equation}
where $F(u,v)$ represents the frequency-domain representation of the image obtained via the Discrete Fourier Transform (DFT) \citep{oppenheim1999discrete}. 
\begin{equation}
    F(u,v) = \sum_{x=1}^{N_{x}^{I}} \sum_{y=1}^{N_{y}^{I}} f(x,y) \cdot e^{-i2\pi \big(\frac{ux}{N_{x}^{I}} +\frac{vy}{N_{y}^{I}} \big)}
\end{equation}
where $f (x, y)$ is the pixel value in the spatial domain, and $(u,v)$ represents the coordinate of a spatial
frequency in the frequency spectrum. Here $e$ is Euler's number, and $i$ is the imaginary unit.

\subsubsection{Evaluation Metrics}
We use three evaluation matrices in this work, namely: max relative loss (MaxRel), Zero Mean Normalized Cross-Correlation (ZNCC) and Structural Similarity Index (SSIM). MaxRel quantifies the relative error, ZNCC measures pixel-wise similarity, and SSIM evaluates perceptual similarity in terms of contrast and luminance.\\

\noindent \textbf{MaxRel(\%)}: This metric quantifies the relative error by comparing the absolute difference between target and emulated images, normalized by the maximum target intensity at each spatial location across the frequency dimension. This normalization ensures that the error is expressed relative to the local peak intensity, providing a scale-independent measure of accuracy.
\begin{equation}
    \text{Median} \Bigg( \frac{1}{n_{\nu}N_{x}^{I} N_{y}^{I}} \sum_{v=1}^{n_{\nu}} \sum_{x=1}^{N_{x}^{I}} \sum_{y=1}^{N_{y}^{I}} 
    \frac{|I_{\nu}(x,y) - I'_{\nu}(x,y)|}{\max_v I_{\nu}(x,y)} \cdot 100\% \Bigg)
    \label{eq:maxrel}
\end{equation}
where $n_{\nu}$ is the number of frequency bins, and \(\max_{\nu} I_{\nu}(x,y) \) is the maximum target intensity at spatial locations \((x,y)\) across all frequencies. \\
Given that \acrshort{coemunet} is intended for application to observational data, and considering the typical scale of observational uncertainties, MaxRel values below 10\% are deemed to indicate sufficient precision in the network's output.

\noindent \textbf{Zero Mean Normalized Cross-Correlation (ZNCC)} is a widely used evaluation metric for template matching~\citep{krattenthaler1994point}, measuring the similarity between two images (signals). It is particularly robust to variations in brightness and contrast between the compared entities. The normalization inherent in ZNCC eliminates contrast differences, while subtracting the mean value removes brightness offsets, ensuring that uniform changes in brightness across an image do not affect the similarity score \citep{di2005zncc} The ZNCC score ranges from $-1$ to $+1$, where $+1$ indicates perfect positive correlation, while $-1$ represents perfect negative correlation, and $0$ suggests no correlation. 

Given two images $I_{\nu}(x,y)$ and $I'_{\nu}(x,y)$,ZNCC is defined as:
\begin{equation}
\text{ZNCC}(I_{\nu}, I'_{\nu}) = \frac{ \sum^{N_{x}^{I}}_{x=1} \sum^{N_{y}^{I}}_{y=1} \left( I_{\nu}(x, y) - \mu_{I_{\nu}} \right) \left( I'_{\nu}(x, y) - \mu_{I'_{\nu}} \right)}{ \sigma_{I_{\nu}} \sigma_{I'_{\nu}}},
\end{equation}
where $\mu_{I_{\nu}}$ and $\mu_{I'_{\nu}}$ are the means of image $I_{\nu}(x, y)$ and $I'_{\nu}(x,y)$, respectively, 
and $\sigma_{I_{\nu}}, \sigma_{I'_{\nu}}$ are their corresponding standard deviations. The final ZNCC score is obtained by averaging across $n_{\nu}$ frequency bins, 
\begin{equation}
    ZNCC = \frac{1}{n_{\nu}} \sum_{\nu=1}^{n_{\nu}} ZNCC(I_{\nu}, I'_{\nu})
\end{equation}

\noindent \textbf{Structural Similarity Index (SSIM)} The SSIM metric \citep{zhou_ssim} quantifies the similarity between two images by considering perceptual factors such as luminance, contrast, and structure. It ranges from -1 to 1, where 1 indicates perfect similarity, 0 represents no similarity, and -1 denotes perfect anti-correlation.

Given two images $I_{\nu}(x,y)$ and $I'_{\nu}(x,y)$, SSIM is defined as:
\begin{equation}
    SSIM(I_{\nu}, I'_{\nu}) = l(I_{\nu},I'_{\nu}) \cdot c(I_{\nu},I'_{\nu}) \cdot s(I_{\nu},I'_{\nu}),
\end{equation}
where the individual components are given by:
\begin{align}
    l(I_{\nu},I'_{\nu}) &= \frac{2 \mu_{I_{\nu}} \mu_{I'_{\nu}}+c_{1}}{{\mu_{I_{\nu}}}^2 +{\mu_{I'_{\nu}}}^2 +c_{1} },\\
    c(I_{\nu},I'_{\nu}) &= \frac{2 \sigma_{I_{\nu}I'_{\nu}} +c_{2}}{{\sigma_{I_{\nu}}}^2 +{\sigma_{I'_{\nu}}}^2 +c_{2} },\\
    s(I_{\nu},I'_{\nu}) &= \frac{\sigma_{I_{\nu}y}+c_{3}}{\sigma_{I_{\nu}} \sigma_{I'_{\nu}}+c_{3} }.
\end{align}
Here $\mu_{I_{\nu}}$ and $\mu_{I'_{\nu}}$ are the mean pixel intensities of $I_{\nu}$ and $I'_{\nu}$, respectively, while $\sigma_{I_{\nu}}$ and $\sigma_{I'_{\nu}}$ are their standard deviations. The term $\sigma_{I_{\nu}I'_{\nu}}$ is the covariance between the two images. The constants $c_1,c_2,c_3$ stabilize the division and are defined as:
\[
c_{1}  = (k_1 L)^2 ,\
c_{2}  = (k_2 L)^2 ,\
c_3    = \frac{c_2}{2},
\]
where $L$ the dynamic range of the pixel-values, and $k_1$ and $k_2$ are small constants to prevent division by zero. The formula for SSIM can be expressed as:
\begin{equation}
SSIM(I_{\nu}, I'_{\nu})=\frac{(2\mu_x \mu_y +c_1)(2\sigma_{xy} + c_2)}{(\mu_x^2+\mu_y^2+c_1)(\sigma_x^2+\sigma_y^2+c_2)}\
\end{equation}
The final SSIM score is computed as the mean SSIM value across all $n_{\nu}$ frequencies, 
\begin{equation}
    SSIM = \frac{1}{n_{\nu}} \sum_{\nu=1}^{n_{\nu}} SSIM(I_{\nu}, I'_{\nu})
\end{equation}

\section{RESULTS}\label{sec:results}

\subsection{Characterizing COEmuNet’s performance}
\begin{figure*}
\includegraphics[width=0.9\linewidth,height=0.6\columnwidth]{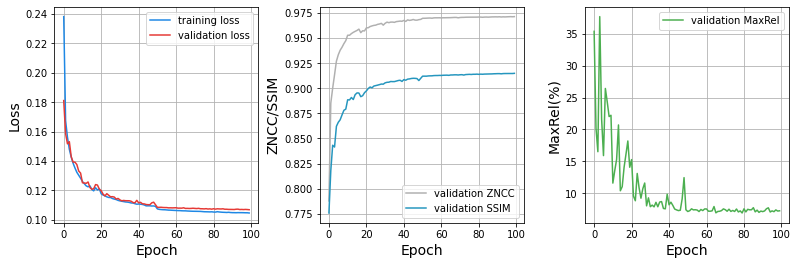}
    \caption{Training history of \gls{coemunet} in terms of the combined frequency (Equation \ref{eq:l1f}) and MSE (Equation \ref{eq:MSE}) loss function (left panel), along with validation metrics (middle and right panels) on validation set. The left panel shows the steady decrease in the training and validation losses, which stabilize around epoch 50, with the training loss converging to approximately 0.11. The middle and right panels present the validation metrics: ZNCC, SSIM, and MaxRel. MaxRel stabilizes below 0.1 after early fluctuations, ZNCC approaches near-optimal values (>0.95), and SSIM converges to high values (>0.9), indicating robust image reconstruction performance. }
    \label{fig:history}
\end{figure*}
\begin{figure}
    \centering
    \includegraphics[width=0.8\columnwidth]{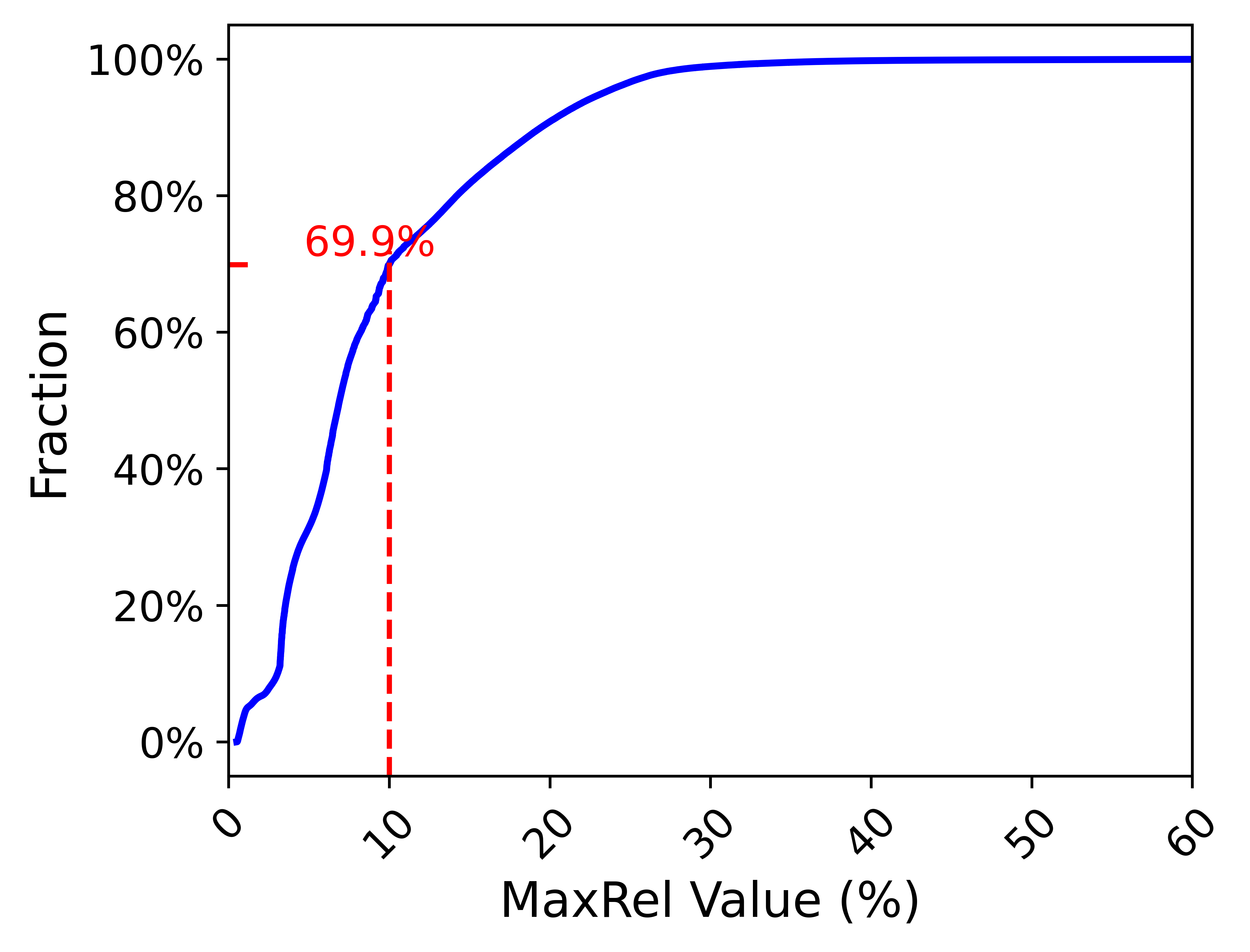}
    \caption{Cumulative distribution of MaxRel values across the test dataset. The steep rise for MaxRel less than 10\% indicates that the majority (69.9\%) of the test dataset has low MaxRel values, while the flatness of the curve beyond 30\% shows that only a small fraction of samples exhibit high errors. This distribution highlights the overall accuracy of \acrshort{coemunet} while also indicating the presence of some outliers with significantly larger errors.}
    \label{fig:maxrel_distribution}
\end{figure}
\begin{table}
\centering
\caption{Evaluation metrics for the test dataset, where lower values of MaxRel ($\downarrow$) and values of ZNCC ($\rightarrow 1$) and SSIM ($\rightarrow 1$) closer to 1 signify better results.}
\label{tab:results_eval}
\begin{tabular}{lrrrr}\toprule
&MaxRel ($\downarrow$) &ZNCC ($\rightarrow 1$) &SSIM ($\rightarrow 1$)\\
\midrule
Test dataset & 6.90\% &0.97 & 0.91 \\
\bottomrule
\end{tabular}
\end{table}
The training history of \acrshort{coemunet} is shown in Figure~\ref{fig:history}. The left-hand subplot shows the training and validation losses, both of which exhibit a steady decline throughout the training period. After approximately 50 epochs, the losses stabilize and show minimal improvement thereafter (a 3.1\% reduction in validation loss), and the training loss converges to around 0.11. 
The small fluctuation observed near epoch 50 is attributed to the resumption of training from a checkpoint (the so-called cold start), which introduces temporary instability in weight calculations, persisting for several epochs. The right-hand subplot shows the evolution of MaxRel, ZNCC and SSIM on the validation dataset over the course of the training. MaxRel exhibits considerable fluctuations during early training but stabilizes below 0.1 after approximately 40 epochs. ZNCC improves rapidly in the initial epochs, approaching near-optimal values ($>0.95$) as training progresses. Similarly, SSIM increases sharply in the early stages and gradually converges to stable high values (>0.9), indicating consistent image reconstruction performance. Given the stabilization of all metrics and marginal improvement beyond 40 to 50 epochs, early stopping is generally recommended at this point to prevent overfitting \citep{Prechelt2012}.  Based on Figure \ref{fig:history}, the training time could be reduced by about 40\% (i.e. stopping after 60 epochs) without compromising the quality of the network performance.

As shown in Table~\ref{tab:results_eval}, \gls{coemunet} achieves a MaxRel of 6.9\%, indicating a low relative error within an acceptable range. The ZNCC score of 0.97, being close to 1, suggests a strong positive correlation between the reconstructed and target images. Additionally, the SSIM value of 0.91 reflects a high degree of structural similarity, demonstrating that the model effectively preserves essential image features during reconstruction.

Figure \ref{fig:maxrel_distribution} presents the cumulative distribution of MaxRel values across the validation dataset. The mean MaxRel value is 9.1\% with a standard deviation of 7\%. The figure shows that 69.9\% of the datasets have MaxRel values below 10\%. It also shows that 20.9\% (8.1\%) of the datasets have MaxRel values above 20\% (30\%). Beyond 30\%, the curve flattens, indicating that only a small fraction of samples exhibit high MaxRel errors. The tail of high MaxRel values indicates a few outliers where the model struggles to capture fine details accurately, such as areas with high complexity or sharp gradients.

\begin{figure*}
    \centering
    \includegraphics[width=0.9\linewidth]{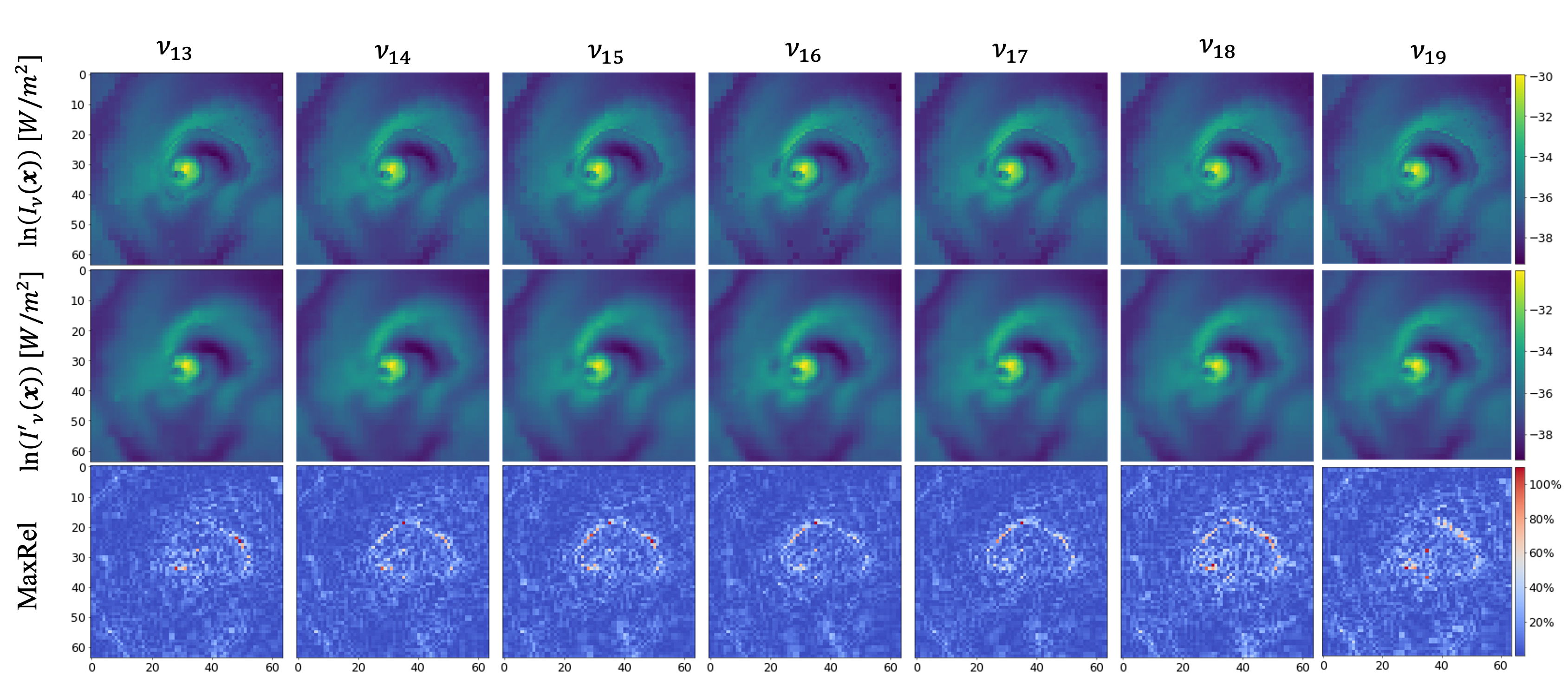}
    \caption{Visualization of intensity maps across multiple frequencies. The top row shows the target intensity maps from frequency $\nu_{13}$ to $\nu_{19}$, while the middle row displays corresponding \acrshort{coemunet} reconstructions across the same frequency range. The bottom row shows the MaxRel values for each pixel in the image. The median MaxRel value across all 7 frequencies is 6.9\% indicating relatively low reconstruction error. The predicted and target images align closely, with minimal differences visible in the MaxRel distribution.}
    \label{fig:mulfreq1}
\end{figure*}

Figure \ref{fig:mulfreq1} presents examples of ground truth images alongside intensity images predicted by \gls{coemunet}. It corresponds to a low MaxRel value of 8.53\%, indicating high reconstruction accuracy. The predicted images closely align with the target images across all frequencies, with minimal deviation observed in the MaxRel error distribution. The most noticeable differences are concentrated in the central regions of the image, where intensity gradients are higher. However, these discrepancies are relatively minor and do not significantly affect the overall reconstruction quality. 

\begin{figure*}
    \centering
    \includegraphics[width=0.9\linewidth]{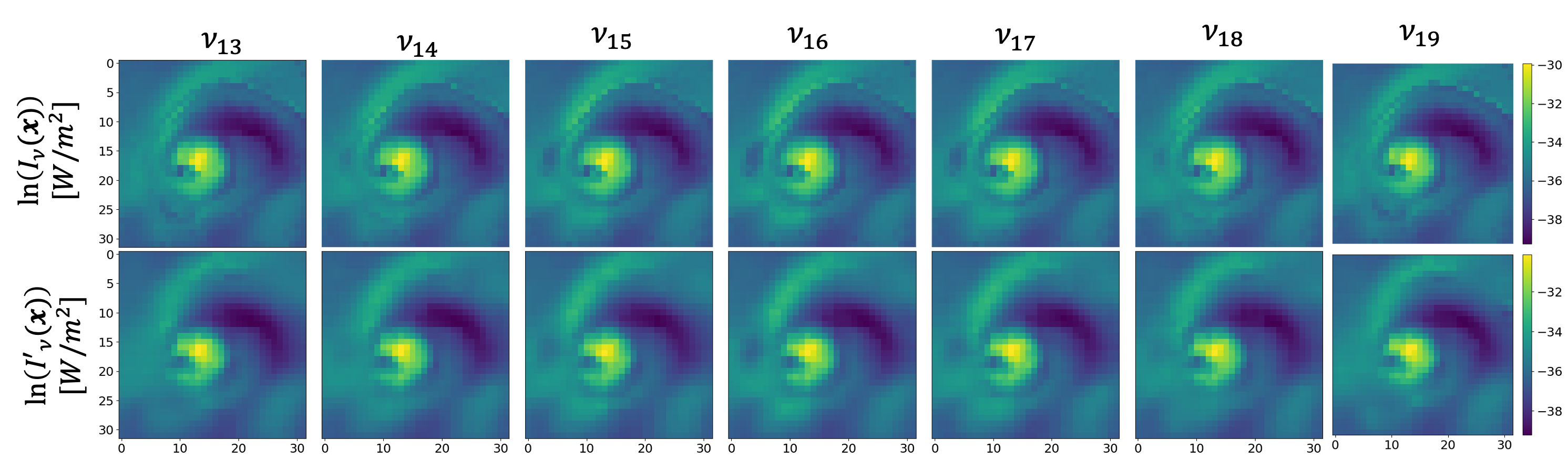}
    \caption{Zoom-in of central 32 by 32 pixels of the target and predicted images from Figure \ref{fig:mulfreq1} illustrating the ability of \acrshort{coemunet} to reproduce detailed features in the target images. See text for further discussion.}
    \label{fig:zoom_in}
\end{figure*}

Figure \ref{fig:zoom_in} presents a zoomed-in view of the central $32 \times 32$ pixel region of the target and predicted images. This highlights the ability of \gls{coemunet} to capture fine features in the images, for example the low density region centred on (10,18) and the large low density region running from ($10,10$) to ($25,20$).

\begin{figure}
    \centering
    \includegraphics[width=0.95\columnwidth]{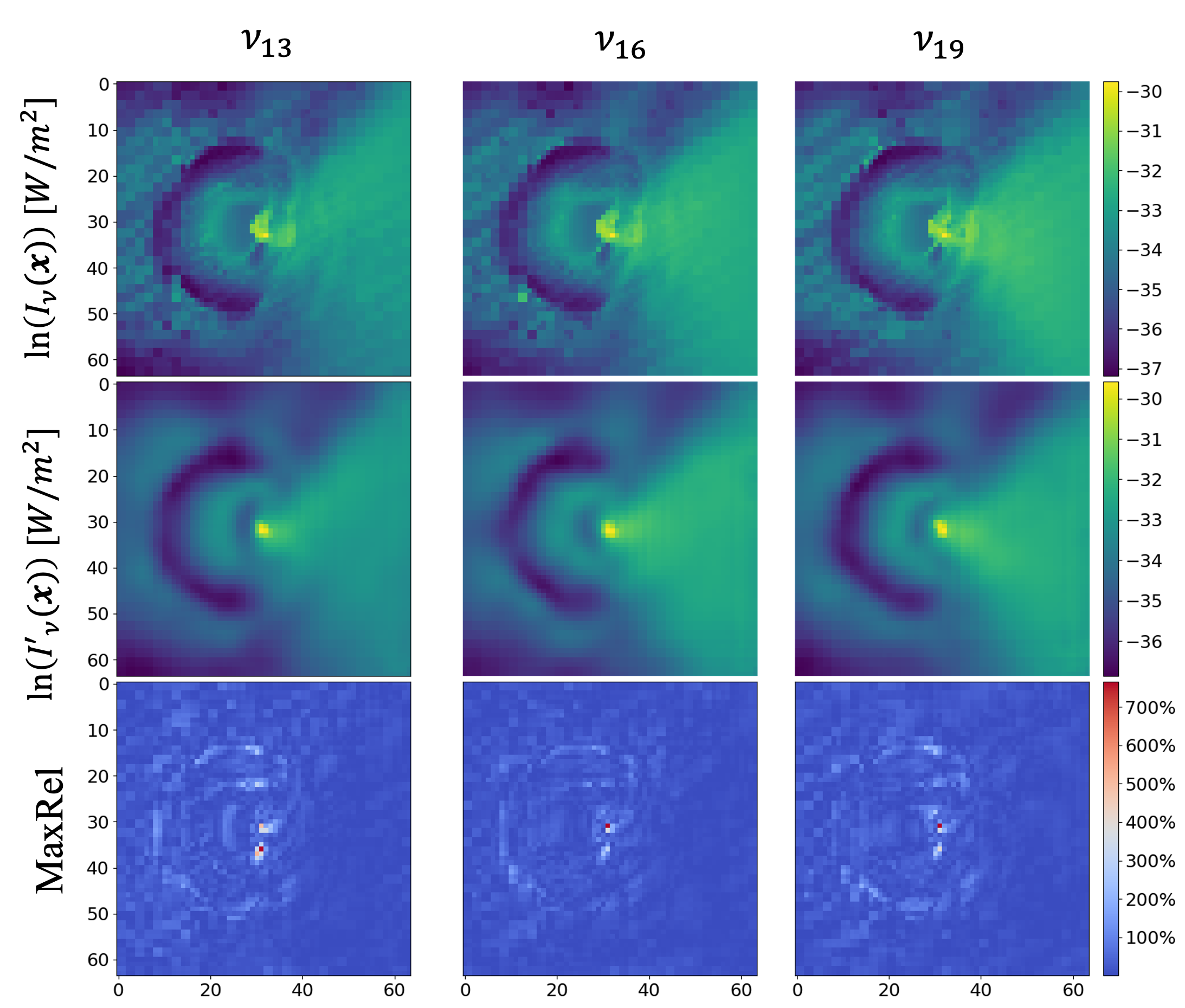}
    \caption{Target and predicted image pairs for 3 frequencies as in Figure \ref{fig:mulfreq1} but for a sample with a Maxrel of 20.94\%. The predicted images generally exhibit moderate errors, apart from in the complex central regions. }
    \label{fig:mulfreq2}
\end{figure}

Figure \ref{fig:mulfreq2} presents target and predicted pairs with a MaxRel value of 20.94\%, indicating a moderate level of reconstruction errors. While the predicted images capture the overall structure of the target images, discrepancies are more pronounced in the complex central regions. 

\begin{figure}
    \centering
    \includegraphics[width=0.95\columnwidth]{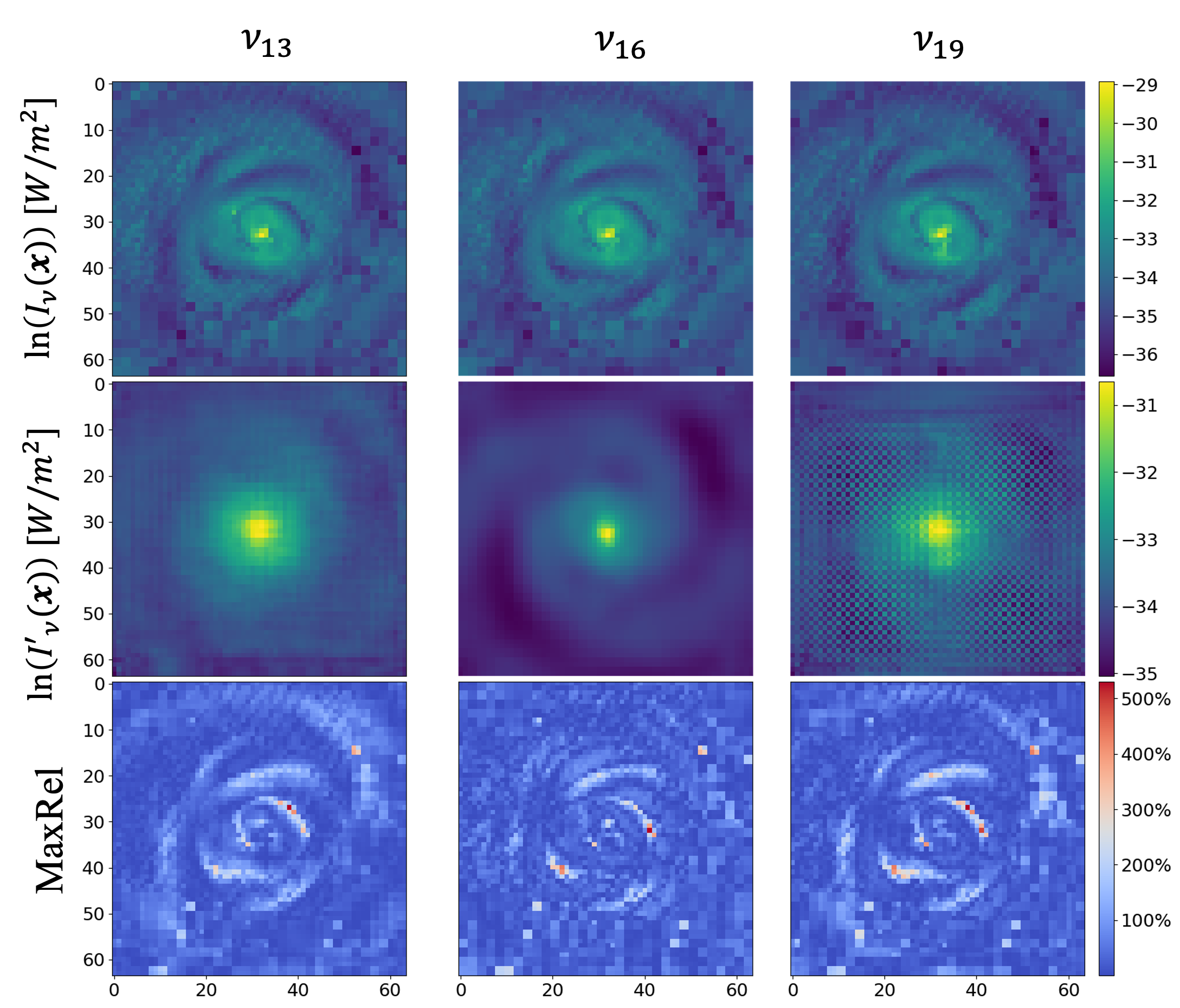}
    \caption{Target and predicted pairs for the validation sample with the highest MaxRel value of 53.10\%, revealing significant reconstruction discrepancies, particularly in the spiral arms and high-intensity regions.}
    \label{fig:mulfreq3}
\end{figure}

Figure \ref{fig:mulfreq3} shows cases with the highest MaxRel values, indicating instances where the model struggles to fully capture the underlying physical processes at this resolution. In a future paper, we will explore a potential route to improving such cases by means of Physics Informed Machine Learning (PIML).

\subsection{Computation speed at different frequencies} 
In this section, we compare the inference times of the trained \gls{coemunet} with the computation times of the Pomme numerical solver, for image data at a resolution of $64 \times 64$. The device in use is an A100-80GB GPU and an Intel Xeon Platinum 8368Q CPU.

Inference time per sample is computed by averaging the batch forward pass time divided by batch size, with mean and standard deviation calculated across batches.

 The quantitative results are presented in Table \ref{tab:runtime_evaluation}. For single frequency (\(n_{\nu}=1\)) datasets $\mathcal{D}'$, \gls{coemunet} running on a GPU achieves a speed-up of $142\times$ compared to Pomme. For multi-frequency (\(n_{\nu}=7\)) dataset $\mathcal{D}$, the acceleration is even more pronounced, with \gls{coemunet} (GPU) being $1007\times$ faster than Pomme.

 Notably, the CPU inference time of \gls{coemunet} is slower than the Pomme numerical slover. This discrepancy arises because neural network inference involves numerous tensor operations, which are optimized for parallel execution on GPUs but are less efficient on CPUs~\citep{Sze2017EfficientPO} In contrast, traditional numerical solvers often leverage vectorized operations, resulting in better performance on CPU architectures.
 
Additionally, \gls{coemunet} exhibits similar inference times for both single- and multi- frequency cases, whereas Pomme's computation time increases by nearly an order of magnitude for the multi-frequency dataset. Overall, the \gls{coemunet} surrogate model significantly outperforms the numerical solver in terms of computational efficiency on both CPU and GPU.
\begin{table*}
\centering
\setlength{\tabcolsep}{12pt}
\caption{Comparison of inference times for \acrshort{coemunet} with computation time for the Pomme numerical solver. The first row presents the results of the single frequency dataset \(\mathcal{D}'\) , while the second row shows the results for the multi-frequency data \(\mathcal{D}\). The table reports the average inference time per sample for \gls{coemunet} using an A100-80GB GPU and an Intel Xeon Platinum 8368Q CPU, as well as the average computation time per sample for the Pomme solver on  an Intel Xeon Platinum 8368Q CPU. See text for detailed discussion.}
\label{tab:runtime_evaluation}
\begin{tabularx}{0.6\linewidth}{@{}l>{\raggedright}Xcc@{}}
\toprule
\shortstack{\textbf{Dataset} \\ \textbf{(64 × 64)}} &
\multicolumn{2}{c}{\shortstack{\textbf{COEmuNet} \\ \textbf{Inference Time (ms)}}} & \shortstack{\textbf{Pomme} \\ \textbf{Calculation Time (ms)}} \\
\cmidrule(lr){2-3}
                 & \textbf{GPU} & \textbf{CPU} & \textbf{CPU} \\
\midrule 

\textit{\(\mathcal{D}'\)(Single-frequency)}  & $0.12 \pm 0.00$  & $61.6 \pm 1.8$ & $17.08\pm0.56$ \\
\textit{\(\mathcal{D}\) (Multi-frequency)}  & $0.13 \pm 0.01$ & $63.9 \pm 2.2$ & $131.54 \pm 3.69$ \\
\bottomrule
\end{tabularx}
\end{table*}

\section{Conclusions}\label{sec:conclusion}
In this work, we presented COEmuNet, an implementation of a 3D Residual Neural Network-based surrogate model for the radiative transfer equation. The proposed model effectively emulates synthetic observations across multiple frequencies, achieving substantial computational speed-up compared to direct numerical computation. The performance of the surrogate model was rigorously evaluated using metrics including Maximum Relative Loss (MaxRel), Zero Mean Normalized Cross-Correlation (ZNCC) and Structural Similarity Index (SSIM).\\

\noindent The results demonstrated a significant computational speed-up, with the model achieving acceleration factors of approximately $1007\times$ for image resolutions of  $64\times64$ on a GPU, compared to the Pomme \citep{de_ceuster_3d_2022} numerical solver. This underscores its potential to facilitate rapid simulations in astrophysical applications. Furthermore, the model achieved a median MaxRel around 6.9\%, a ZNCC above 0.95 and an SSIM exceeding 0.9, demonstrating its strong predictive accuracy. This high fidelity is evident in the model's ability to capture detailed spatial structures and spectral dependencies, closely replicating the ground truth images.\\

\noindent While the proposed model demonstrates strong performance, it has certain limitations. First, the \gls{coemunet} presented here has been trained exclusively on data for a specific spectral line CO($J=2\rightarrow1$) and frequency range ($n_\nu =7$). Secondly, the current model is tailored to binary-perturbed AGB star outflows; its extension to other geometries (e.g., single-star winds or circumstellar disks) has not yet been explored and may require retraining or fine-tuning on domain-specific data.
Thirdly, the \gls{coemunet} does not incorporate physical principles, which limits its full potential in leveraging deep learning capabilities. In a future paper, we will explore the integration of Physics Informed Machine Learning (PIML) to enhance model performance by embedding physical constraints into the learning process. Overall, this research illustrates the potential of deep learning in addressing the computational challenges associated with \acrfull{rt} calculations, paving the way for more efficient and scalable modelling techniques in astrophysics.
\section*{Acknowledgements}
SS is supported by a PhD studentship from the Science and Technology Facilities Council (STFC) and a College of Science and Engineering Scholarship from the University of Leicester. FDC was supported by a Postdoctoral Research Fellowship of the Research Foundation - Flanders (FWO), grant number 1253223N and by a KU Leuven Postdoctoral Mandate (PDM), grant number PDMT2/21/066. The hydrodynamic simulation data were generated as part of ATOMIUM DiRAC project (dp147). The project was also partially funded by the BASE-II ExCALIBUR project (EPSRC grant number EP/X019918/1).\\

This work used the DiRAC Data Intensive service (CSD3) at the University of Cambridge, managed by the University of Cambridge University Information Services on behalf of the STFC DiRAC HPC Facility (\url{www.dirac.ac.uk}). It also used the DiRAC Extreme Scaling service (Tursa) at the University of Edinburgh, managed by the Edinburgh Parallel Computing Centre on behalf of DiRAC. DiRAC servies were funded by BEIS, UKRI and STFC capital funding and STFC operations grants. DiRAC is part of the UKRI Digital Research Infrastructure. 
\section*{Conflict of Interest}
Authors declare no conflict of interest.

\section*{Data Availability}
 The data underlying this article will be shared on reasonable request to the corresponding author. The
code used to produce the results in our study is available in a github repository \url{https://github.com/beaversliv/3D_RTE_surrogate_model}


\newpage
\bibliographystyle{mnras}
\bibliography{example} 


\newpage
\appendix

\section{Supplementary material}
In this Appendix, we describe how we constructed synthetic images of a hydrodynamic model from arbitrary observation directions, further experimental configuration details and robustness results.
\subsection{Rotation of the 3D Hydrodynamic Model}\label{sec:rotate_hydro_model}
\begin{figure}
    \centering
    \includegraphics[width=0.9\linewidth]{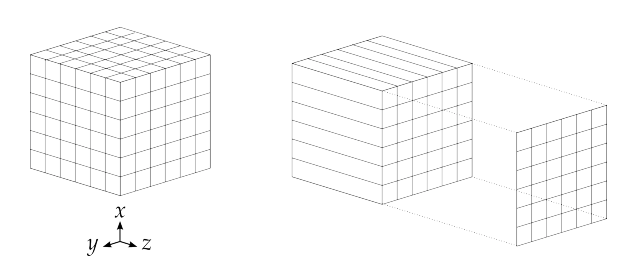}
    \caption{(\textit{Left}.) Graphical representation of a 3D variable as a 3D PyTorch tensor. (\textit{Right}.) Graphical representation of how radiative transfer is performed, by solving the line integrals along the z-axis of the tensor, producing an image in the xy-plane. \citep{de2024bayesian}}
    \label{fig:hydro_grid}
\end{figure}
To analyse hydrodynamic models from arbitrary observing directions, it is necessary to account for the particular constraints of the Pomme~\citep{de2024bayesian} numerical solver, which computes radiative transfer exclusively along the $z-$axis. To generate synthetic observations from different directions, each hydrodynamic model $H_{i}(\boldsymbol{x})\in \mathbb{R}^{3}$ must be reoriented such that the desired viewing direction aligns with the solver’s fixed axis.

The reorientation process begins by trimming $H_{i}(\boldsymbol{x})$ to remove empty regions. $H_{i}(\boldsymbol{x})$ is then rotated using a transformation matrix \( R(x,y,z) \) ,  aligning it with the fixed direction required by the numerical solver. Once reoriented, the intensity\( I_\nu(\mathbf{x}, \boldsymbol{\hat{n}}) \)
is computed along the $z-$ axis.
\subsubsection{Construction of an Auxiliary Spherical Model}
To prevent voids that may arise during rotation, an auxiliary spherical model $H_{s}(\boldsymbol{x})$ is constructed to provide a consistent spatial domain for all hydrodynamic models. These voids occur because rotating a 3D cubic model and re-gridding it in a fixed 3D space can create empty regions between the edges of the rotated model and the reformulated grid. 

The process begins by computing a local radius $r_i$ for each hydrodynamic model $H_{i}(\boldsymbol{x})$, defined as:
\begin{equation*} 
r_i = \min \{ \frac{x_{\max,i} - x_{\min,i}}{2}, \frac{y_{\max,i} - y_{\min,i}}{2}, \frac{z_{\max,i} - z_{\min,i}}{2} \}, 
\end{equation*}
where \(x_{\min,i}\) and \(x_{\max,i}\) represent the minimum and maximum $x$-coordinates, respectively, with analogous definitions for the $y$ and $z$ dimensions. Concurrently, the centroid $c_i$ of each model is computed as:
\begin{equation*}
   c_i = (\frac{x_{max,i}+x_{min,i}}{2},\frac{y_{max,i}+y_{min,i}}{2},\frac{z_{max,i}+z_{min,i}}{2})
\end{equation*}
To define the auxiliary spherical model $H_{s}(\boldsymbol{x})$, a radius $r$ is determined as the smallest local radius across all hydro models:
\begin{equation*}
    r = \min \{ r_1, r_2, \dots, r_i \}, \quad i \in N.
\end{equation*}
Each point \(x_i,y_i,z_i\) in a hydrodynamic model \(H_{i}(\boldsymbol{x})\) is then evaluated for inclusion within $H_{s}(\boldsymbol{x})$. The Euclidean distance $d$ between a point \(x_i,y_i,z_i\) and its centroid \(c_i\) is computed as:
\begin{equation*}
    d = \sqrt{(x_{i,c}-x_i)^2 + (y_{i,c}-y_i)^2 +(z_{i,c}-z_i)^2}
\end{equation*}
Points satisfying the condition $d>r$ are removed, ensuring all hydrodynamic models fit within the global spherical domain.

\subsubsection{Rotation of the Hydrodynamic Model}
Following the trimming process, $H_{i}(\boldsymbol{x})$ is uniformly rotated within the three-dimensional space. Given a model $H_{i}(\boldsymbol{x})$, a rotation matrix $R(x,y,z)$ is applied to transform its coordinates:
\[
\begin{pmatrix}
x' \\
y' \\
z'
\end{pmatrix}
= R(x,y,z)
\begin{pmatrix}
x \\
y \\
z
\end{pmatrix},
\]
\subsection{Experimental Settings}\label{sec:settings}
Table \ref{tab:faceon_pretrain} shows  the configuration of the network used during pre-training of the COEmuNet and Table\ref{tab:mulfreq_finetune} is the configuration used for fine-tuning the COEmuNet.
\begin{table}
    \caption{Pretraining settings of $I_{\nu}(\hat{\boldsymbol{n}},\boldsymbol{x}), n_\nu = 1$}
    \begin{tabular}{l|l}
    \toprule
    \textbf{Pre-training config} &   \\
    \midrule
    Loss function & $\lambda=0.8, 1-\lambda=0.2$  \\ 
    optimizer& Adam  \\ 
    base learning rate& $4e-3$  \\ 
    weight decay & 0\\
    optimizer momentum & $\beta_1, \beta_2=0.9,0.999$ \\
    batch size & 512 \\
    training epochs & 1000\\
    learning rate scheduler & StepLR \\
    step size & 200 \\
    decay rate & 0.1 \\
    \bottomrule
    \end{tabular}
    \label{tab:faceon_pretrain}
\end{table}

\begin{table}
    \caption{Fine tune settings}
    \begin{tabular}{l|l}
    \toprule
    \textbf{Fine-tuning config} &   \\
    \midrule
    Loss function & $\lambda=0.8, 1-\lambda=0.2$  \\  
    optimizer & Adam  \\ 
    base learning rate & 0.0001  \\ 
    weight decay & 0\\
    optimizer momentum & $\beta_1, \beta_2=0.9,0.999$ \\
    batch size & 2048 \\
    training epochs & 100 \\
    learning rate scheduler & None\\
    \bottomrule
    \end{tabular}
    \label{tab:mulfreq_finetune}
\end{table}

\subsection{Robustness Evaluation}
To assess the stability and generalization capability of the trained models, we conducted a robustness evaluation by training \acrshort{coemunet} with different random seeds. Table \ref{tab:random_seed_with_zncc} presents a low variance in all three evaluation metrics across different seeds, indicating that the model is robust and stable.
\begin{table}
    \caption{Robustness test on $I_{\nu}(z,\boldsymbol{\hat{n}})$ where $n_{\nu} = 1, \boldsymbol{\hat{n}}=(0,1)$}
    \begin{tabular}{@{}cccc@{}}
    \toprule
    \textbf{Random seed} & \textbf{MaxRel} $\downarrow$ &\textbf{ZNCC} $\uparrow$ & \textbf{SSIM} $\uparrow$\\
    \midrule
    1234& 6.8001\% & 0.9856 & 0.9677 \\ 
    42 & 7.0455\%  & 0.9880 & 0.9712\\ 
    0 & 6.3684\%   & 0.9876 & 0.9698\\
    12345 & 6.5940\% & 0.9852 & 0.9660 \\
    \bottomrule
    \label{tab:random_seed_with_zncc}
    \end{tabular}
\end{table}

\end{document}